# Suppression of the inclination instability in the trans-Neptunian Solar system

Arnav Das [1]★ and Konstantin Batygin[2]

[1]*Division of Physics, Mathematics and Astronomy, California Institute of Technology, Pasadena, CA 91125, USA*
[2]*Division of Geological and Planetary Sciences, California Institute of Technology, Pasadena, CA 91125, USA*

## ABSTRACT

The trans-Neptunian scattered disc exhibits unexpected dynamical structure, ranging from an extended dispersion of perihelion distance to a clustered distribution in orbital angles. Self-gravitational modulation of the scattered disc has been suggested in the literature as an alternative mechanism to Planet nine for sculpting the orbital architecture of the trans-Neptunian region. The numerics of this hypothesis have hitherto been limited to $N < O(10^3)$ superparticle simulations that omit direct gravitational perturbations from the giant planets and instead model them as an orbit-averaged (quadrupolar) potential, through an enhanced $J_2$ moment of the central body. For sufficiently massive discs, such simulations reveal the onset of collective dynamical behaviour – termed the 'inclination instability' – wherein orbital circularisation occurs at the expense of coherent excitation of the inclination. Here, we report $N = O(10^4)$ GPU-accelerated simulations of a self-gravitating scattered disc (across a range of disc masses spanning 5–40 $M_\oplus$) that self-consistently account for intraparticle interactions as well as Neptune's perturbations. Our numerical experiments show that even under the most favourable conditions, the inclination instability never ensues. Instead, due to scattering, the disc depletes. While our calculations show that a transient lopsided structure can emerge within the first few hundreds of Myr, the terminal outcomes of these calculations systematically reveal a scattered disc that is free of any orbital clustering. We conclude thus that the inclination instability mechanism is an inadequate explanation of the observed architecture of the Solar system.

## 1 INTRODUCTION

Objects in the outer reaches of our system exhibit a range of observational peculiarities, from their clustering in angular and physical space to their fortuitously high perihelion distances, alongside an accompanying gap in observed perihelia. Moreover, a fraction of the so-called extreme trans-Neptunian objects (eTNOs) exhibits dramatically high inclinations from the ecliptic (for a review, see Batygin et al. 2019).

A multitude of explanations exists for the aforementioned orbital anomalies. First, there persists the notion that the above irregularities are a figment of survey strategy and thereby do not supply any physical implications. Whilst this posited null hypothesis has engendered a healthy amount of contention in the past (Shankman et al. 2017; Bernardinelli et al. 2020; Napier et al. 2021), any isolated analysis of observational biases pertinent to individual surveys cannot confidently conclude physical clustering of these objects, largely due to the limited sky coverage of the surveys themselves. On the other hand, comprehensive considerations (through combined observability analyses of all available data) have yielded clustering in the Runge–Lenz and angular momentum vectors of distant Kuiper Belt objects (KBOs) at a 99.6 per cent significance level (Brown

2017; Brown & Batygin 2019, 2021). Further, the existence of high-inclination orbits and the perihelion gap categorically escapes the null hypothesis, and we note from Kaib et al. (2019) that demotions of Oort cloud objects into the Kuiper belt facilitated by Galactic perturbations does not provide an influx of sufficiently inclined KBOs to match the observed distribution. All this inevitably serves to suggest that our explanation should take the form of a steady external effect that likely continues to gravitationally sculpt the trans-Neptunian architecture of our Solar system.

Several such mechanisms have been probed in the literature: planets heralding von Zeipel–Lidov–Kozai dynamics (de la Fuente Marcos & de la Fuente Marcos 2014; Trujillo & Sheppard 2014), shepherding of objects by a distant, lopsided massive planetesimal disc (Sefilian & Touma 2019), which is envisioned to potentially manifest through the so-called 'inclination instability' (Madigan & McCourt 2016; Zderic & Madigan 2020), and, finally, the Planet nine hypothesis (Batygin & Brown 2016a; Batygin et al. 2019). Batygin & Brown (2016a) have shown that Kozai–Lidov models demand the unorthodox presence of multiple distant planets in close proximity to the individual TNOs, which even then would cluster only the arguments ($\omega$) and not the longitudes of perihelion ($\varpi$) and would fail to generate confinement in the orbital poles (i.e. inclination, $i$, and longitude of ascending node, $\Omega$). Likewise, although the self-gravity of long-term stable, eccentric shepherding discs does lead to the confinement of $\varpi$ through an apsidally anti-aligned secular equilibrium at high eccentricities (Sefilian & Touma 2019), that such a delicately sustained orbital configuration would survive the history of the Solar system is questionable.

★ E-mail: arnav257@caltech.edu

On the contrary, the influence of a singular distant, eccentric, super-Earth-type object – Planet nine – that is apsidally anti-aligned with the observed KBOs furnishes both an analytically and numerically satisfactory account of all the aforementioned anomalies. Through distinct octupole-level dynamics, Planet nine simultaneously modulates the degrees of freedom related to the TNO's eccentricities, inclinations, nodes, and apsidal angles in a way that leads to the observed $\varpi$ confinement and the orbit-flipping behaviour seen in high-inclination KBOs (Batygin et al. 2019). This model also explains the detachment of distant TNOs from Neptune's orbit, and the observed perihelion gap has been found withal in subsequent numerics within this framework (e.g. Oldroyd & Trujillo 2021). Other verifiable abductive predictions of the Planet nine paradigm further bolster its likelihood: observations of TNO orbits that are roughly perpendicular to the ecliptic can be demonstrated appreciably in simulations with Planet nine (Batygin & Morbidelli 2017); intricate interplay with Neptune not only depopulates the objects aligned with Planet nine's orbit but also populates the trans-Jovian system with high-inclination centaurs (Batygin & Brown 2016b; Batygin & Morbidelli 2017; Becker et al. 2018), which have been observed in considerable proportion (Gomes, Soares & Brasser 2015).[1]

Notwithstanding these predictions merited by the above theoretical framework, the aforementioned 'inclination instability' mechanism actuated in primordial scattered discs of icy planetesimals has been purported as a potent counter-hypothesis in the literature. In a string of analyses (Madigan & McCourt 2016; Madigan et al. 2018a; Zderic & Madigan 2020; Zderic et al. 2020, 2021), the authors find that the self-gravity of highly eccentric, multi-$M_{\oplus}$ debris discs triggers a global exponential growth in inclination, a concomitant decrease in eccentricity, and a secular broadening of perihelia. Through the emergence of a lopsided orbital configuration (Madigan et al. 2018a; Zderic et al. 2020), the desired inclination excitations, perihelion dispersions, and $\varpi$ clustering (Zderic et al. 2021) are promised to emerge in our Solar system. In Zderic & Madigan (2020), the authors further argue that a sufficiently high primordial disc mass can resist the gas and ice giant-induced differential precession that seeks to wash out the inclination instability (utilising quadrupolar harmonics to emulate their Gaussian averaged influence). In effect, Zderic & Madigan (2020) propose the existence of a $\sim$20 $M_{\oplus}$ (Earth mass) primordial scattered disc, of which the observed KBOs would be a staggeringly low proportion. In this work, we test the veracity of this picture.

Modelling the long-term evolution of the primordial scattered disc has posed the lasting challenge of computational expense, with previous studies having to approximate either self-gravity (e.g. Levison et al. 2011) or giant planet influence (Zderic & Madigan 2020; Zderic et al. 2020, 2021) through computationally inexpensive roundabouts. The latter contributions neglect the exact resolution of the outer planets (including only their secular averages instead). In modelling a realistic history of the primordial disc, however, omission of short-term scattering effects induced by these planets is self-inconsistent – in that the modulation provided by Neptune-scattering is assumed to have given rise to the scattered disc structure in the first place, yet these short-term effects are neglected in modelling subsequent disc dynamics. Here, we alleviate these limitations.

In this paper, we present direct, self-consistent *N*-body calculations for a suite of primordial discs to show that the inclination instability forecasted in the literature fails to effectuate under the influence of Neptune's full gravitational perturbations. In Section 2, we motivate our methods and describe the simulations conducted, with enhanced resolution in the number of particle interactions modelled and the integration time-scales. In bootstrapping the underlying numerics with GPU acceleration, we permit the computational capacity to simulate realistic disc masses, self-gravity, as well as Neptune's orbit. In Section 3, we explicate our results and compare them with prior predictions, highlighting the role of apsidal precession and Neptune-scattering in suppressing self-gravitational modulation in the Solar system. Lastly, in Section 4, we discuss our findings, reason through their implications, and outline how this work refutes the inclination instability as a viable explanation for the observations at hand.

## 2 METHODS

### 2.1 Numerical modelling

To date, numerical models of self-gravity within the scattered disc have neglected the consequences of close encounters with Neptune. Particularly, in Zderic & Madigan (2020) and Zderic et al. (2020, 2021), the authors account solely for the phase-averaged influence of the giant outer planets in their vindication of a quasi-stable, 20 $M_{\oplus}$ trans-Neptunian scattered disc. The process of Neptune scattering, however, is well-known to eject objects from the Solar system – and deplete the disc thereby. Reducing Neptune self-inconsistently to its corresponding effective quadrupolar moment thus qualifies as a notable operational deficiency.

A second limitation of the existing calculations is that they are restricted to $N \leqslant O(10^3)$ superparticles and low simulation time-scales. To circumvent this issue, the aforementioned studies invoke unrealistically high disc masses of $\sim$10$^{-3}$ $M_{\odot}$ to mimic the dynamical history of the Solar system (Zderic & Madigan 2020). Although inclination instability by pure self-gravity has been shown to yield interesting dynamics in such *hypothetical* orbital configurations, rigorous investigation of this mechanism in the trans-Neptunian Solar system necessitates higher numerical resolution. In Zderic et al. (2021), the authors rely on heuristic scaling arguments (involving disc masses, effective $J_2$ moments, and instability saturation time-scales) to translate their results to the Solar system and procure $\sim$20 $M_{\oplus}$ as the requisite disc mass-scale for the inclination instability to ensue.

We obviate the need to engage with this scheme (and other aforementioned approximations) by utilising GPU-accelerated *N*-body simulations. Using the second-order hybrid symplectic integrator code, GENGA (Grimm & Stadel 2014), we are empowered with the computational speedup to simulate physical $M_{\odot}$: $M_{\text{planets}}$: $M_{\text{disc}}$ ratios as well as resolve Neptune's orbit self-consistently in computing close encounters, which GENGA handles using Bulirsch–Stoer changeovers. Further, we simulate the scattered disc as $N = O(10^4)$ fully interacting objects over multi-Gyr time-scales, putting these models on par with standard numerical explorations in the Planet nine paradigm (Batygin et al. 2019; Batygin & Brown 2021; Brown & Batygin 2021). For completeness, we also enforce the formal convergence requirement of any multipolar expansions, such that objects falling inside the effective radius of the utilised quadrupolar harmonics are removed from the computations. We note that preliminary work done in Emel'yanenko (2022) includes exact resolution of the giant planet orbits over comparable time-scales (for a different variant of the scattered disc), but these simulations account for only $\sim O(10^2)$ intra-disc superparticle interactions, approximating self-gravity and inclination instability in a disc of otherwise massless test particles; nevertheless, we remark that this work corroborates the robustness of the rapid depletion that we demonstrate in Section 3.

---

[1] Such dynamics could even extend to retrograde Jupiter Trojans (Köhne & Batygin 2020).

### 2.2 Initial orbital parameters

We distil the ingredients of the setup in Zderic & Madigan (2020) and Zderic et al. (2021) to simulate fully interacting scattered discs in the outer Solar system. Following the aforementioned studies, we initialise the disc masses to be $O(20\,\mathrm{M_\oplus})$, which, in absence of planetary perturbations, is sufficient to exhibit inclination instabilities on Gyr time-scales (see Fig. 1). Following suit further, we initialise the orbits of our disc with fixed perihelion values of $q_0 = 30$ au and uniformly distributed semimajor axes of $a_0 \in (100$ and $1000$ au), yielding eccentricities of $e_0 \in (0.7, 0.97)$. We stick with these values of $q_0$ and $a_0$ throughout the paper, with the exception of Section 3.1 and Appendix A, where we initialise $q_0 \in (30$ and $36$ au) and draw semimajor axes from an $a^{-1.5}$ distribution over the same range instead (obtaining essentially identical results). As the disc is initially axisymmetric, we draw initial mean anomalies $\mathcal{M}_0$, arguments of perihelion $\omega_0$, longitudes of ascending node $\Omega_0$, and thereby longitudes of perihelion $\varpi_0$, uniformly from $(0°, 360°)$. Lastly, we draw inclinations $i_0$ from Rayleigh distributions around mean values of $5°$. We simulate such discs with total primordial masses of $0\,\mathrm{M_\oplus}$, $5\,\mathrm{M_\oplus}$, $10\,\mathrm{M_\oplus}$, $20\,\mathrm{M_\oplus}$, $30\,\mathrm{M_\oplus}$, and $40\,\mathrm{M_\oplus}$. For each case, different permutations of the orbital configurations are modelled, and characteristic dynamics are here reported. To visualise orbits, we also use the utility modules of the REBOUND software package (Rein & Liu 2012).

### 2.3 Simulating the giant planets

To resolve Neptune in our numerics, we initialise its present-day orbital elements, verifying that orbital migration of the planet due to scattering is negligible over 5 Gyr runs. To account for the phase-averaged influence of the remaining giant outer planets (Jupiter, Uranus, and Saturn), we add a zonal quadrupolar harmonic $J_2$ to the central body of the system as per,

$$J_2 = \frac{1}{2\,\mathrm{M_\odot} R^2} \sum_{i=1}^{N} m_i a_i^2 \tag{1}$$

with effective radii $R = 25$ au across all simulations with Neptune.

In simulations with Neptune, $N = 3$ (corresponding to the three giant outer planets) and $J_2 \approx 5 \times 10^{-5}$. Likewise, in simulations where Neptune's scattering is omitted (so that the planet is phase-averaged instead), $N = 4$ and $R = 13$ au, so that $J_2 \approx 3 \times 10^{-4}$.

We remove all objects that fall within the effective radius $R = 25$ au (physically interpreting this inner truncation as entrapment into the centaur and eventually the Jupiter-family comet system), with the exception of illustrative simulations that omit self-consistent planetary perturbations altogether – in such cases, we inner truncate using $R_\odot \approx 0.005$ au (Fig. 1) and $R = 13$ au (Fig. 2) instead. Similarly, we set the outer truncation radius in all simulations as $10^4$ au, a separation beyond which perturbations from Galactic tide become appreciable. In agreement with the 25 au boundary imposed above, we limit ourselves to a time-step of $\delta t \approx 10$ yr ($\sim$1/16th of Neptune's orbital period).

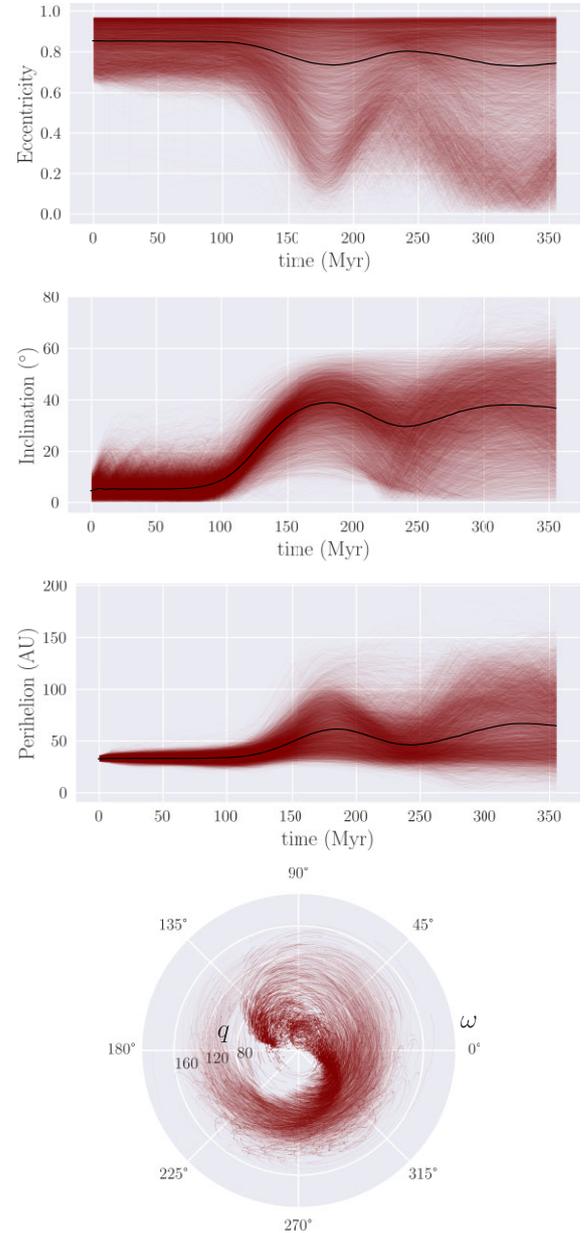

**Figure 1.** Inclination instability in a $20\,\mathrm{M_\oplus}$ self-gravitating disc (omitting all planetary perturbations). Individual orbits (maroon) and their collective medians (black) are shown (left). Global orbital circularisation is captured in the eccentricity decrease (first panel), at the expense of coherent inclination excitation (second panel) and secular perihelion dispersion (third panel). Argument of perihelion ($\omega$) clustering is presented through a trace of objects in a polar $\omega$–$q$ plane (fourth panel) over the last 25 Myr of this simulation.

In summary, we simulate a primordial scattered disc comprising tens of $M_\oplus$ initialised with the orbital conditions outlined in Section 2.2, emulating the secular influence of Jupiter, Saturn, and Uranus through their corresponding Gaussian-averaged $J_2$ from equation (1). Inversely, we model Neptune's gravity directly to account for scattering. The bulk of this contribution thus serves to adjudicate the inclination instability proposal upon full remediation of all the known dominant drivers in the dynamics of the trans-Neptunian Solar system.

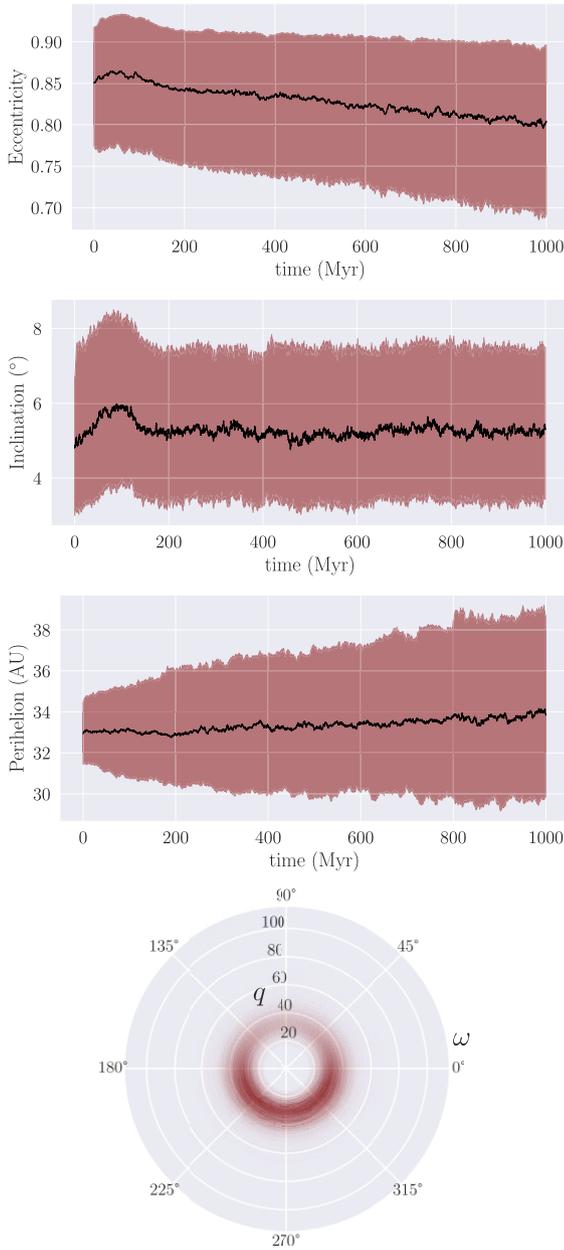

**Figure 2.** Orbital evolution for a $20\,M_\oplus$ scattered disc with added $J_2$ (emulating all giant outer planets). Medians are denoted in black for eccentricity (first panel), inclination (second panel), and perihelion (third panel). Regions between the upper and lower quartiles are, respectively, shaded. In the fourth panel, traces for the final 200 Myr are shown in a polar $\omega$–$q$ ($^\circ$–au) plane. In contrast with Fig. 1, $\omega$ clustering and $e$ and $i$ oscillation are absent. Inclination instability is suppressed by the secular influence of the giant planets.

## 3  RESULTS

### 3.1  Preliminaries

In Madigan et al. (2018a), the authors express caution for using *N*-body simulations to study collective self-gravity phenomena, arguing that the strength of secular effects is generally suppressed – when compared to theory – for finite (especially low) *N*. In simulating the self-gravitational evolution of the scattered disc, however, we find quantitative agreement on instability time-scales over a range of *N* upon systematically scaling up to $O(10^4)$, which strongly indicates convergence in modelling the relevant secular dynamics.

Using the GPU-accelerated numerics outlined in Grimm & Stadel (2014), we first model a $20\,M_\oplus$ variant of the '*sd100*' configuration in Zderic & Madigan (2020) in the absence of any external perturbers (with the initial conditions outlined in Section 2.2). In Zderic & Madigan (2020), the surface densities utilised to derive the $20\,M_\oplus$ threshold is $\Sigma \propto a^{-2}$. We find that for discs with steeper surface density profiles, the inclination instability manifests on shorter time-scales. Here, we use a steeper $\mathrm{d}N/\mathrm{d}a \propto a^{-1.5}$ semimajor axis distribution as this corresponds to the steady-state surface density of $\Sigma \propto a^{-2.5}$, which follows from a Fokker–Planck treatment of Neptune-scattering as a diffusive process in specific energy. Importantly, when disc depletion through planetary perturbations is accounted for, we find that the initial surface density profile makes no difference to the interplay of emergent effects (this is seen through a comparison of Section 3.2 and Appendix A). Further, for our initial orbital configuration, we use $q_0 \in (30 \text{ and } 36 \text{ au})$ to moderate Neptune-stirring when it is added at $\sim$30 au to this system (see Appendix A). This setup highlights a clear onset of the inclination instability that activates in a disc under its pure self-gravity (see Fig. 1). Coherent eccentricity decrease, inclination upsurge, and perihelion growth are accompanied by a broadening of their initial confinements (see Section 2.2) to the terminal spreads captured in Fig. 1. Quasi-oscillatory changes in the medians of each quantity are observed on time-scales of $O(100\,\mathrm{Myr})$. Notably, orbits can reach dramatically high inclinations, including those with retrograde configurations. We highlight argument of perihelion ($\omega$) confinement through the trace of a cluster of objects seen to sweep clockwise through $\omega$–$q$ space (shown for the last 25 Myr in Fig. 1; fourth panel). All objects stay within the boundary of this setup throughout the reported time interval. These observations are in compliance with standard results in the inclination instability literature. We find, however, that relative to $\omega$, there is no statistically significant longitude of perihelion ($\varpi$) clustering in the terminal distributions at $t = 1$ Gyr.

To build complexity sequentially, we add to the above system a quadrupolar deformation in the central potential corresponding to the $J_2$ of all four outer planets (from equation 1) with an effective radius of $R = 13$ au. For computational convenience, we use $N = O(10^3)$ in modelling this disc up to 1 Gyr. It is hypothesised in Zderic et al. (2021) that a $20\,M_\oplus$ primordial scattered disc with this $J_2$ should sustain intermittent apsidal clustering up to $O(\mathrm{Gyr})$ time-scales and exhibit the requisite inclination instabilities saturating at $\sim$250 Myr. We find instead that the inclination instability is inhibited in a primordial disc of $20\,M_\oplus$ even within this idealised setup, suggesting the existence of confounders that break the qualitative scaling arguments that yield such values. This is due, in part, to a steady outward depletion of the disc mass effected by the inclusion of this $J_2$.

In Fig. 2, we report this inhibition through the distinct eccentricity and inclination signatures – through the median, upper, and lower quartile evolution of the disc – over 1 Gyr and a 0.8–1 Gyr trace in $\omega$–$q$ space (which fails to exhibit the form of clustering seen in Fig. 1). We draw particular attention to the orbital inclinations, which plateau at $i \lesssim 10^\circ$ in the first Gyr itself (despite the observed perihelion dispersion, which occurs instead in the absence of inclination excitation; see Section 3.3). Further, in Madigan et al. (2018a), Zderic & Madigan (2020), and Zderic et al. (2021), the characteristic inclination instability time-scale is taken to be inversely proportional to disc mass. Because $J_2$ is scale-dependent, however, we find these scaling arguments to be valid only in cases where self-gravity is modelled in the absence of planetary perturbations. In simulations with $J_2$, the steady disc mass depletion caused by objects dispersing beyond the $R = 10^4$ au boundary also plays a secondary role in inflating the inclination instability time-scale (in some analogy to the Neptune-stirring simulations described below). In effect, the precession induced by the giant planets is itself sufficient to inhibit the inclination instability in our $20\,\mathrm{M}_\oplus$ scattered disc configuration.

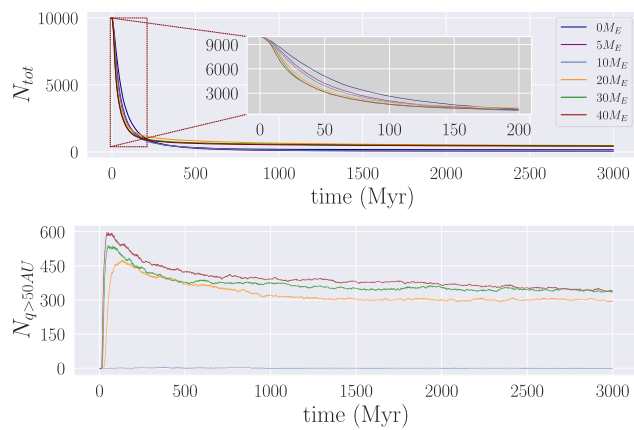

**Figure 3.** Rapid disc depletion effectuated by Neptune's scattering as a function of primordial mass (top). More massive discs succeed more readily in dispersing perihelia (shown for an arbitrary boundary of 50 au; bottom).

### 3.2 Realistic disc evolution

Although intriguing, the above systems are fundamentally unphysical given the omission of short-term Neptune-scattering in the simulations. Therefore, for the remainder of this section, we compare the long-term dynamics of various primordial disc masses upon resolving Neptune exactly (and phase-averaging the other giant planets). To contrast a counterfactual sample, we also provide the evolution of a massless disc of test particles initialised identically.

The addition of Neptune scattering leads to an almost immediate ejection of an overwhelming majority of objects from the system, acting further against the inclination instability in unison with the apsidal precession driven by the added $J_2$. This leads to total inhibition of the inclination instability mechanism. In Fig. 3, we highlight the rapid depletion through which Neptune dominates across all disc masses for a disc with a $a^{-1}$ surface density profile. The terminal distributions are limited to $N \approx O(10^2)$ objects (with a meagre $\lesssim 5$ per cent maximal residue of the primordial mass), with higher-mass primordial discs retaining a higher proportion of planetesimals. Moreover, under the constraints $40\,\mathrm{au} < q < 100\,\mathrm{au}$ and $250\,\mathrm{au} < a < 2000\,\mathrm{au}$, which filters for the pertinent observational census of trans-Neptunian orbits, only primordial discs with $\geq 20\,\mathrm{M}_\oplus$ show noteworthy residuals. In Fig. 4, we present one such typical remnant in contrast with the observed collection of orbits in the outer Solar system. As noted, the differences are most starkly observed through the absence of any apsidal clustering produced by our models. At any rate, we find the characteristic time-scale for rapid depletion here to be significantly lower than that of the inclination instability, in agreement with past predictions (Nesvorný 2015; Batygin & Brown 2016a). In effect, rapid depletion disbars the scattered disc from undergoing the inclination instability in the Solar system.

In Fig. 5, we report the evolution for eccentricity, inclination, and perihelion of objects in a $20\,\mathrm{M}_\oplus$ scattered disc with the inclusion of Neptune-scattering, presenting also the terminal distributions of pertinent orbits at $t = 5\,\mathrm{Gyr}$ for each quantity. Across the entire simulation suite, it is noted that any characteristic signature of the inclination instability is fully inhibited upon inclusion of short-term scattering for all primordial discs (see Appendix A for a disc with a $a^{-2.5}$ surface density profile and Appendix C for other masses). In particular, we highlight that the inclinations plateau between ∼10–15° for all masses, in contrast with the ∼50° upsurge seen in the pure self-gravity case in Fig. 1.

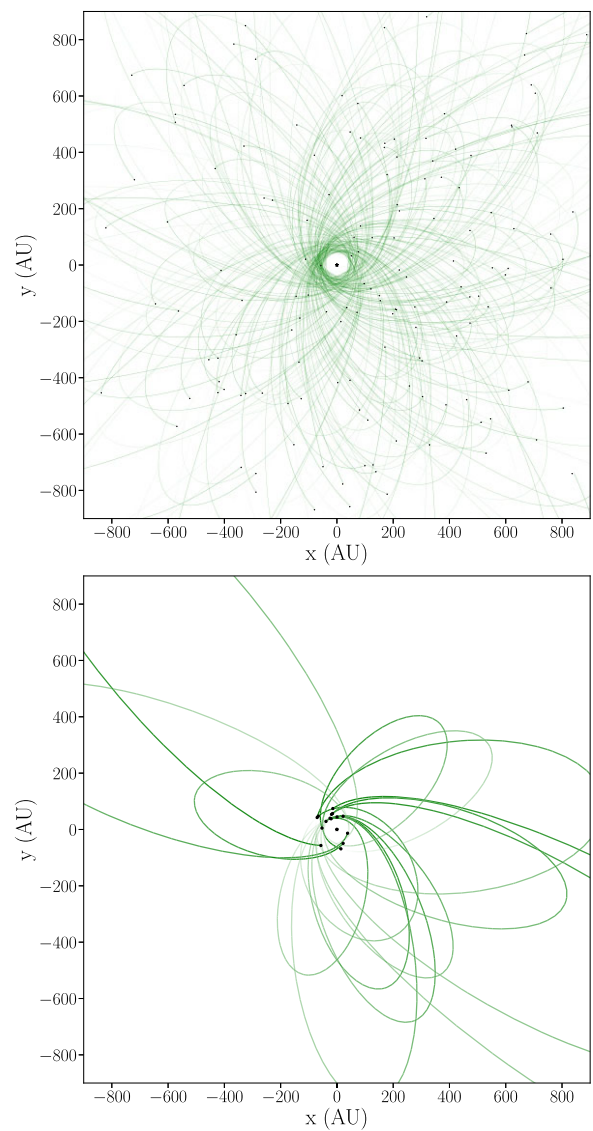

**Figure 4.** Characteristic residual orbital configuration for $O(20\,\mathrm{M}_\oplus)$ primordial scattered discs (top) and observational data (bottom). All orbits are subject to the constraints $40\,\mathrm{au} < q < 100\,\mathrm{au}$ and $250\,\mathrm{au} < a < 2000\,\mathrm{au}$.

As is expected given the suppression of the inclination instability, apsidal clustering in the residual population of $N \sim O(10^2)$ planetesimals (for $\geq 20\,\mathrm{M}_\oplus$ discs) is absent. In Fig. 6, we present the terminal distributions of orbital poles in a $20\,\mathrm{M}_\oplus$ disc through the final Gyr traces of observable orbits in polar $\varpi$–$q$, $\omega$–$q$, and $\Omega$–$i$ planes. Across all disc masses (including $<20\,\mathrm{M}_\oplus$), no statistically significant clustering is found in final $\varpi$ or $\omega$, and no asymmetric grouping is observed in $\Omega$–$i$ values, highlighting the clear absence of any terminal apsidal clustering in this model. In our simulations, we find that the competition between two $O(10\,\mathrm{Myr})$ effects – Neptune-scattering and self-shepherding – in the higher mass discs incidentally leads to a transient one-off asymmetry in apsidal angles at $O(100\,\mathrm{Myr})$ time-scales (see Appendix B), wherein orbits stochastically pick a direction in $\varpi$–$q$ space to disperse outwards, through a mode of perihelia broadening disparate from the inclination instability (see Section 3.3): though unrelated to the Solar system, these dynamics are of some academic interest. At any rate, however, this transient apsidal asymmetry is swiftly replaced by an isotropic distribution in apsidal angles over all perihelia bins satisfying 40 au $< q < 100$ au. This uniformity endures over $O$(Gyr) time-scales to yield the final Gyr footprints of Fig. 6.

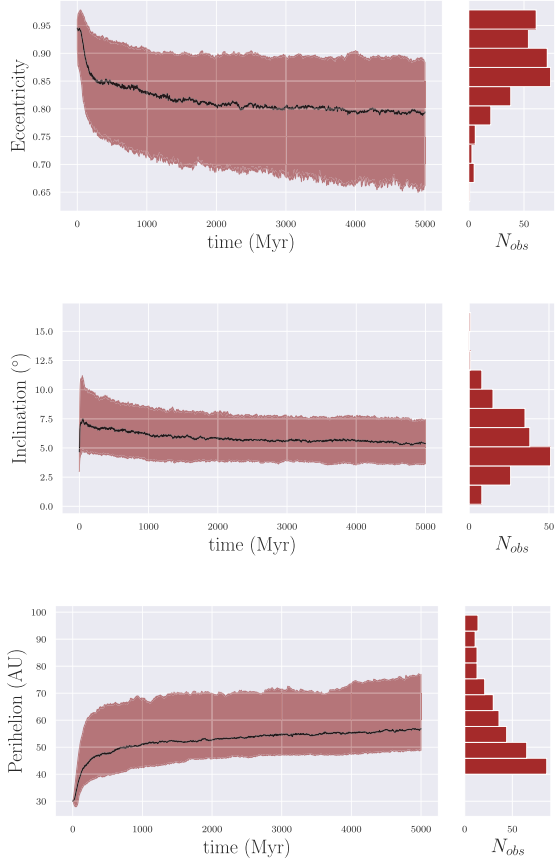

**Figure 5.** Long-term evolution of a 20 $M_\oplus$ scattered disc with self-consistent resolution of Neptune (and Jupiter, Uranus, and Saturn as $J_2$). Medians are denoted in black for eccentricity (first panel), inclination (second panel), and perihelion (third panel). The regions between the upper and lower quartiles for each quantity are shaded, and terminal distributions at $t = 5$ Gyr are provided on the right (when filtered for the observational region). In contrast with Figs 1 and 2, no quasi-oscillatory signatures of the inclination instability are observed, with inclinations plateauing below $\sim 15°$. Inclination instability is thus observed to be fully suppressed by Neptune-scattering.

### 3.3 Secular perihelion broadening

Although any signatures of the inclination instability are absent in our simulations that account for Neptune self-consistently, we find a completely distinct mode of secular perihelion growth that affects a small proportion of objects through self-modulation in more massive discs.

A clear instantiation of this self-modulation is captured in the terminal $a$–$q$ distribution of Fig. 7 for a 40 $M_\oplus$ disc. Importantly, we find that the peaks of these distributions, irrespective of their modalities, consistently inhabit regions of $a$–$q$ space (for all disc masses $\geq 20 M_\oplus$) where prior results in the literature, based on the instability manifesting, have evinced a vacated perihelion gap

(Zderic & Madigan 2020; Zderic et al. 2020). As shown in Fig. 7, we find the contrary to be true. This further spurs the inference that self-gravity alone is an incomplete story.

The observation of perihelion (and eccentricity) broadening hints at parallel dynamics discovered in Madigan et al. (2018b), whereby in-plane gravitational torque by a lopsided disc on individual orbits is found to be responsible for individual secular oscillations in eccentricity without any inclination excitation, as we observe in Fig. 5. With a more massive primordial reservoir, more accentuated perihelion dispersion, as well as a higher depletion rate, is seen. This suggests that prior to ejection, in-plane gravitational torque (which would be higher for a more massive disc) initiates secular dynamics in the eccentricities of individual orbits, ultimately resulting in their reaching the boundaries of the simulation (primarily the inner boundary) and further depleting the disc. In effect, as Neptune-stirring rapidly modulates the disc on time-scales of $O$(10 Myr), self-gravity broadens the perihelia on comparable time-scales (following the aforementioned dynamics). We find that as a consequence of this competition, only discs with masses $\geq 20 M_\oplus$ are able to shepherd objects to perihelia of $O$(100 au) through their self-modulation. Self-gravity – through this chain of interactions – empowers a small proportion of objects to escape to high perihelia, resulting in a minor population of planetesimals permanently insulated from close encounters with Neptune; the higher the mass of the primordial reservoir, the higher is the proportion of this isolated population. In Fig. 3 (and through Figs 5 and C2), we make explicit how higher mass discs emplace a larger proportion of planetesimals to an arbitrarily chosen boundary. In any case, however, there emerges no apsidal clustering within the observable regions of $q \lesssim 100$ au in these calculations.

Our numerical experiments have thus demonstrated that upon the self-consistent inclusion of Neptune's short-term scattering, the inclination instability mechanism fails to manifest in the outer Solar system due (in large part) to rapid orbital depletion. Collective gravity in $\gtrsim O$(20 $M_\oplus$) discs permits sufficient self-modulation to shepherd small proportions of orbits to high perihelia, but these displaced remnants of highly massive primordial reservoirs possess neither the inclination spread, nor the $\varpi$ clustering, nor the perihelion gap that is seen to exist in the distant Solar system. In theory and in practice, the inclination instability mechanism in primordial scattered discs fails to explain the observational peculiarities that persist in the trans-Neptunian Solar system.

## 4 DISCUSSION

We began this paper by outlining several ideas that have cycled through the literature of trans-Neptunian orbital dynamics (in Section 1). Inclination instability, however, has endured in hinting at possible resolutions for lasting problems. In this work, we have used GPU-accelerated simulations to model the long-term evolution of $O$(20 $M_\oplus$) primordial scattered discs and shown how they fail to exhibit the inclination instability.

The key feature of our simulations that directly leads to the inhibition of the inclination instability is the self-consistent inclusion of Neptune. We highlight how the disc rapidly depletes on characteristic time-scales far shorter than that of the inclination instability, fully suppressing any actuation of the latter effect. Fig. 5 presents the real evolutions of orbital eccentricity and inclination values (when Neptune-stirring is permitted in the disc), with terminal residuals yielding acutely insufficient spreads in either quantity and a complete lack in orbital clustering. Additionally, we demonstrate that the perihelion gap previously claimed to emerge from self-gravity of the primordial disc never realises. Lastly, we show that no signs of $\varpi$ clustering are found in the postdicted observable remnants of this physical setup.

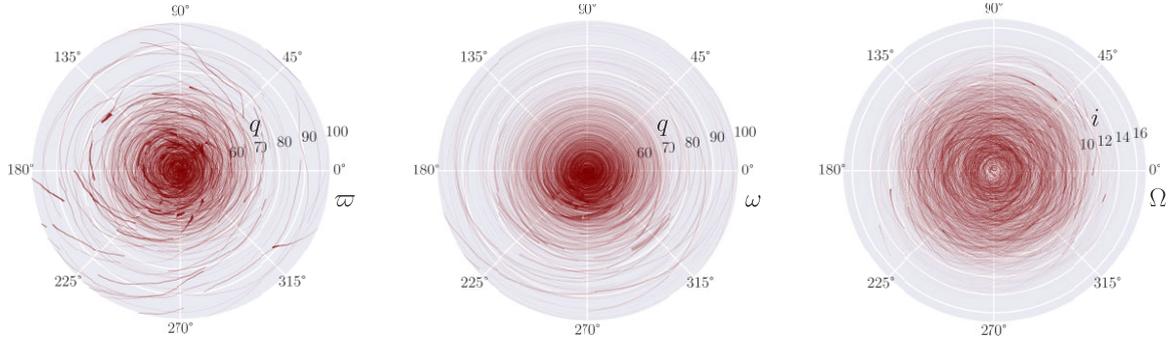

**Figure 6.** Terminal Gyr traces in a polar $\varpi$–$q$, $\omega$–$q$, and $\Omega$–$i$ plane for objects in $20\,\mathrm{M}_\oplus$ primordial scattered discs. As seen in the left-hand and middle panel, no statistically significant clustering is found in $\varpi$ or $\omega$. In the right-hand panel, we observe no grouping in $\Omega$–$i$ space (with $i$ in °), indicating the absence of confinement in orbital poles. This inclination instability model thus fails to postdict apsidal clustering upon the complete inclusion of giant outer planet influence.

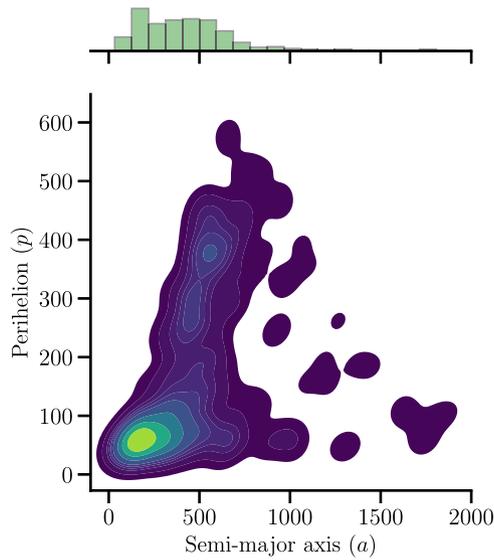

**Figure 7.** Terminal $a$–$q$ distribution for a $40\,\mathrm{M}_\oplus$ disc at $t = 5\,\mathrm{Gyr}$ amidst competition between Neptune-scattering and self-modulation. Marginal histograms are provided for both $a$ and $q$. In massive discs ($\geq 20\,\mathrm{M}_\oplus$), significant perihelion dispersion is observed for a small proportion of objects, leading to terminal perihelia values of $O(100\,\mathrm{au})$. Similar $a$–$q$ distributions are found in 20 and $30\,\mathrm{M}_\oplus$ cases.

The only detectable effects that realise through the collective self-gravity of such discs are reflected in the perihelion broadening (in the yet unobservable $q > 100\,\mathrm{au}$ region) that accentuates with increasing primordial disc mass. Of particular academic interest is the competition between Neptune-scattering and such perihelia dispersion. The higher the primordial mass supply, the more marked the latter effect, with more of the primordial disc being accrued at high enough perihelia to evade Neptune's perturbations. This effect doubtlessly warrants some interest, but it ultimately arises in absence of the requisite apsidal confinement or angular momentum vector clustering that is seen in the data.

The inclination instability models have to date been posited to provide equivalent reconstructions as the Planet nine paradigm. Considerations of close encounters with the giant planets, however, introduce a fundamental disparity in the viability of these two explanations. Whereas depletion of a massive scattered disc is exceedingly likely, especially through interactions with Neptune (as we have shown here), no corresponding short-term effect can regulate the orbital sculpting facilitated by Planet nine as readily. On the contrary, it is the interplay between Planet nine's secular modulation and Neptune-scattering that manifests as the emergence of apsidal clustering within the context of this hypothesis (see Batygin et al. 2019).

In summary, our assessment of the inclination instability has shown that both the orbital precession *and* the short-term scattering induced by the giant outer planets is sufficient to suppress the forecasted self-gravity effects within the scattered disc. Irrespective of what is found truly lurking in the outer Solar system, the inclination instability paradigm has herein been demoted in supplying any clear plausible progress to this end.

## ACKNOWLEDGEMENTS

The computations presented here were conducted primarily in the Resnick High Performance Computing (HPC) Center, a facility supported by Resnick Sustainability Institute at Caltech. The authors acknowledge the David and Lucile Packard Foundation for their generous support.

We are grateful to our reviewer, Dr. Ann-Marie Madigan, for her thoughtful feedback on this work. Special thanks are also extended to Simon L. Grimm for releasing updated GENGA modules better suited for specific application to this work.

## DATA AVAILABILITY

The data underlying this article, consisting primarily of *N*-body simulations carried out using GPU-accelerated code, can be shared upon reasonable request to the corresponding author.

## APPENDIX A: THE EFFECT OF NEPTUNE ON DISCS WITH STEEPER DENSITY PROFILES

As we have shown throughout Section 3, the action of Neptune on a $20\,M_\oplus$ primordial scattered disc leads to the suppression of the inclination instability. These results were obtained with the initial conditions specified in Section 2.2, whereby the initial semimajor axes were uniformly distributed in the range $a_0 \in (100$ and 1000 au). Here, we initialise a more physical $dN/da \propto a^{-1.5}$ distribution in semimajor axis. As in the systems of Figs 1 and 2, we also bootstrap a dispersed perihelion distribution by setting $q_0 \in (30$ and 36 au), primarily to avoid extremising the initial scattering between Neptune and the disc. We find that even with these initial configurations, the inclination instability remains fully inhibited. In Fig. A1, we present the eccentricity, inclination, and perihelion evolution for this system with $N = O(10^4)$ to highlight that the qualitative account remains the same as that outlined in Section 3. In effect, the results of these simulations are robust to changes in the initial surface density profile of the disc.

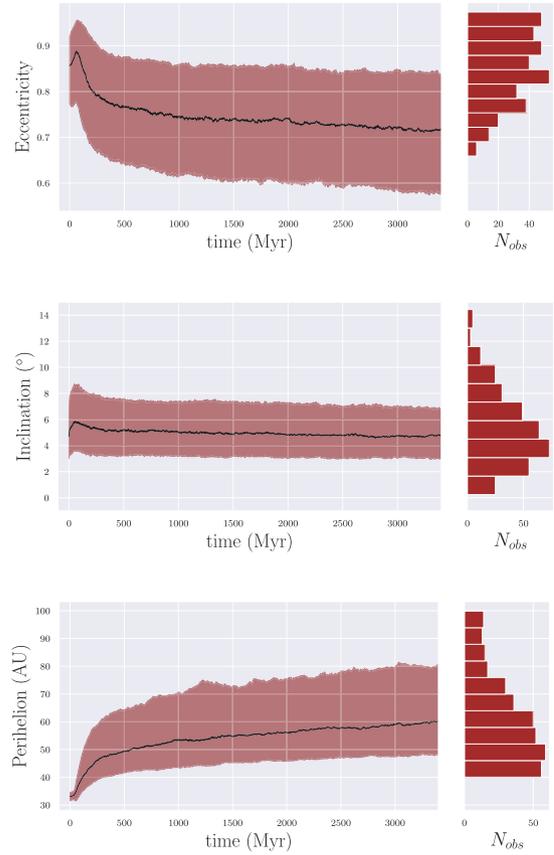

**Figure A1.** Long-term evolution of a $20\,M_\oplus$, $dN/da \propto a^{-1.5}$ scattered disc with self-consistent resolution of Neptune (and Jupiter, Uranus, and Saturn emulated as $J_2$). Medians are denoted in black for eccentricity (first panel), inclination (second panel), and perihelion (third panel). The regions between the upper and lower quartiles for each quantity are shaded, and terminal distributions at $t \sim 3.5$ Gyr for the observational region are presented on the right. As in the $a^{-1}$ surface density case of Fig. 5, no inclination instability is observed.

## APPENDIX B: TRANSIENT $\varpi$ CLUSTERING IN HIGH-MASS INTERMEDIATE DISCS

We have shown, through Figs 4 and 6, that a 5 Gyr remnant of $O(20 \, M_\oplus)$ scattered discs evolves into an axisymmetric uniform $\varpi$ configuration. However, at time-scales of $O(100 \, \text{Myr})$, primordial discs with $\geq 20 \, M_\oplus$ exhibit significant asymmetry in $\varpi - q$ space, as we highlight in Fig. B1 for $t = 0$–500 Myr. Discs with higher masses are shown to self-modulate more readily, shepherding their objects to distant perihelia at a higher rate (in both the distances to which objects are emplaced and the proportion of the primordial reservoir that is emanated effectively; e.g. Fig. 7). This shepherding is shown here to occur through a preferred (stochastic) direction in $\varpi$ space; however, this asymmetry dissipates completely over $O(\text{Gyr})$ time-scales, more rapidly so for higher-mass discs.

This raises the pertinent question of whether it is possible to fine-tune the initial disc masses to yield peak (apparent) apsidal clustering at $t = 5 \, \text{Gyr}$. As we have shown in Fig. 3, discs with $\leq 10 \, M_\oplus$ fail to counter Neptune's disc depletion, making them highly ineffective in shepherding objects to high perihelia in the first place. This cuts off the possibility of sufficiently lowering the initial disc mass to prolong this $\varpi$ asymmetry up to $t \sim 5 \, \text{Gyr}$ time-scales. Within realistic primordial scattered discs, fine-tuning the initial conditions for inclination instability to yield observable apsidal clustering is thus unfeasible and, at any rate, supplies a narrative that is unduly incomplete.

## APPENDIX C: INCLINATION INSTABILITY INHIBITION IN 0, 5, 30, AND 40 $M_\oplus$ DISC MASSES

In Section 3, we present the inhibition of inclination instability resulting from our numerical calculations of a 20 $M_\oplus$ primordial scattered disc. Here, we report similar results for a 0, 5, 30, and 40 $M_\oplus$ disc. Across the simulation suite, no inclination instability is observed to survive. In Figs C1 and C2, we present the evolution of eccentricity, inclination, and perihelia through median, upper, and lower quartile values for each quantity in 0, 5, 30, and 40 $M_\oplus$ discs. As in Section 3, we also include traces for the final Gyr of each object in a polar $\varpi - q$ plane, highlighting the lack of any apsidal clustering in all disc masses.

In all cases, the inclination instability mechanism is suppressed. In cases with disc masses $\lesssim 20 \, M_\oplus$, the disc fails to trigger the secular perihelion dispersion discussed in Section 3.3. This is reflected in Fig. C1 for a 0 and 5 $M_\oplus$ disc, where the supermajority of objects are confined within a 40 au perihelion distance. Vice versa, discs with primordial mass $\gtrsim 20 \, M_\oplus$ are seen to disperse in perihelia, as captured in Fig. C2.

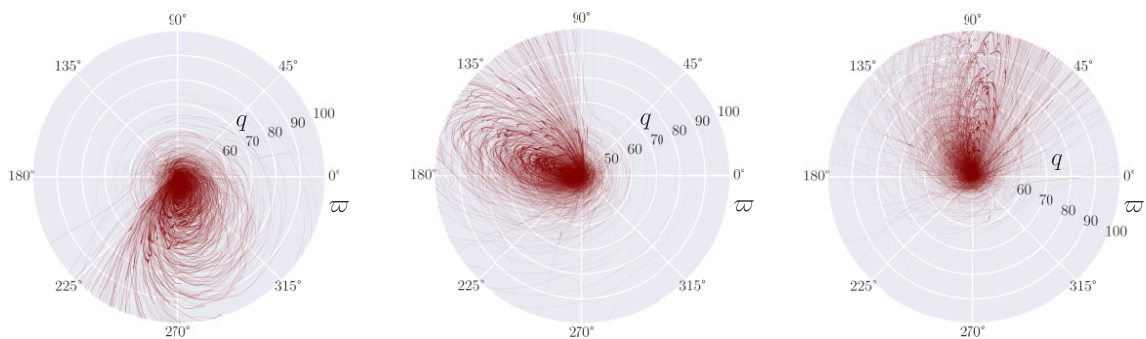

**Figure B1.** $\varpi - q$ footprints for 20, 30 and 40 $M_\oplus$ scattered discs over the first 0.5 Gyr. Significant asymmetry is found in $\varpi$ as objects disperse outwards in $q$ stochastically. This asymmetry is dissipated on integration time-scales of $\sim O(0.5 \, \text{Gyr})$, and a quasi-symmetric distribution in $\varpi$ is attained by $t \approx 2 \, \text{Gyr}$. At $t = 5 \, \text{Gyr}$, this isotropy persists to yield the terminal outcomes presented in Fig. 6.

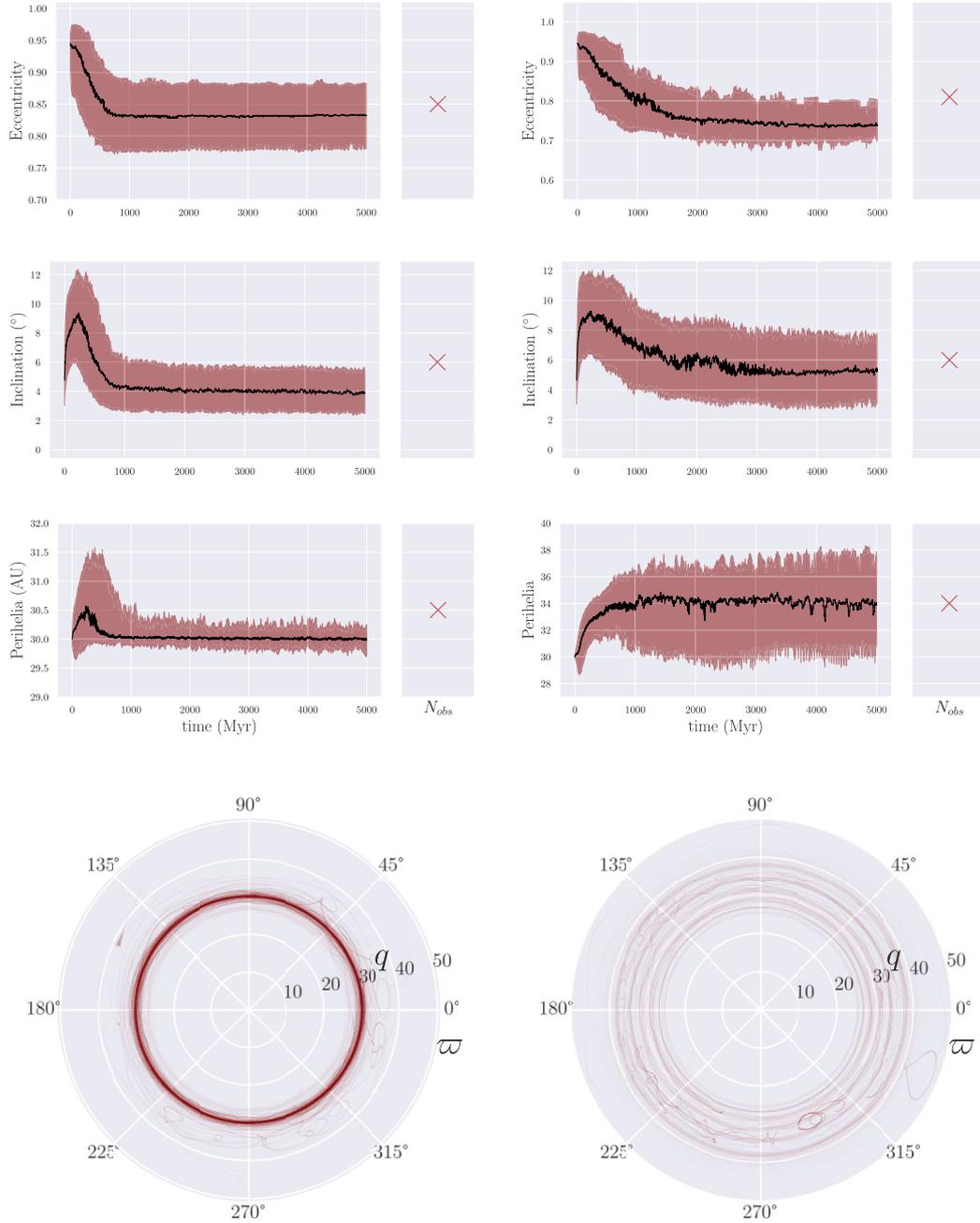

**Figure C1.** Long-term evolution of a $0\,\mathrm{M_\oplus}$ (left column) and $5\,\mathrm{M_\oplus}$ (right column) scattered disc with self-consistent resolution of Neptune (and Jupiter, Uranus, and Saturn modelled as a $J_2$ moment). Medians are denoted in black for eccentricity (first row), inclination (second row), and perihelion (third row). The regions between the upper and lower quartiles for each quantity are shaded, and terminal distributions at $t = 5$ Gyr provided on the right report that no objects are found within the observational region. In the fourth row, traces for the final Gyr are presented in $\varpi$–$q$ space. No secular perihelion dispersion is seen as in the cases of Figs 5 and C2. All characteristic signatures of the inclination instability are inhibited, and no $\varpi$ confinement is found.

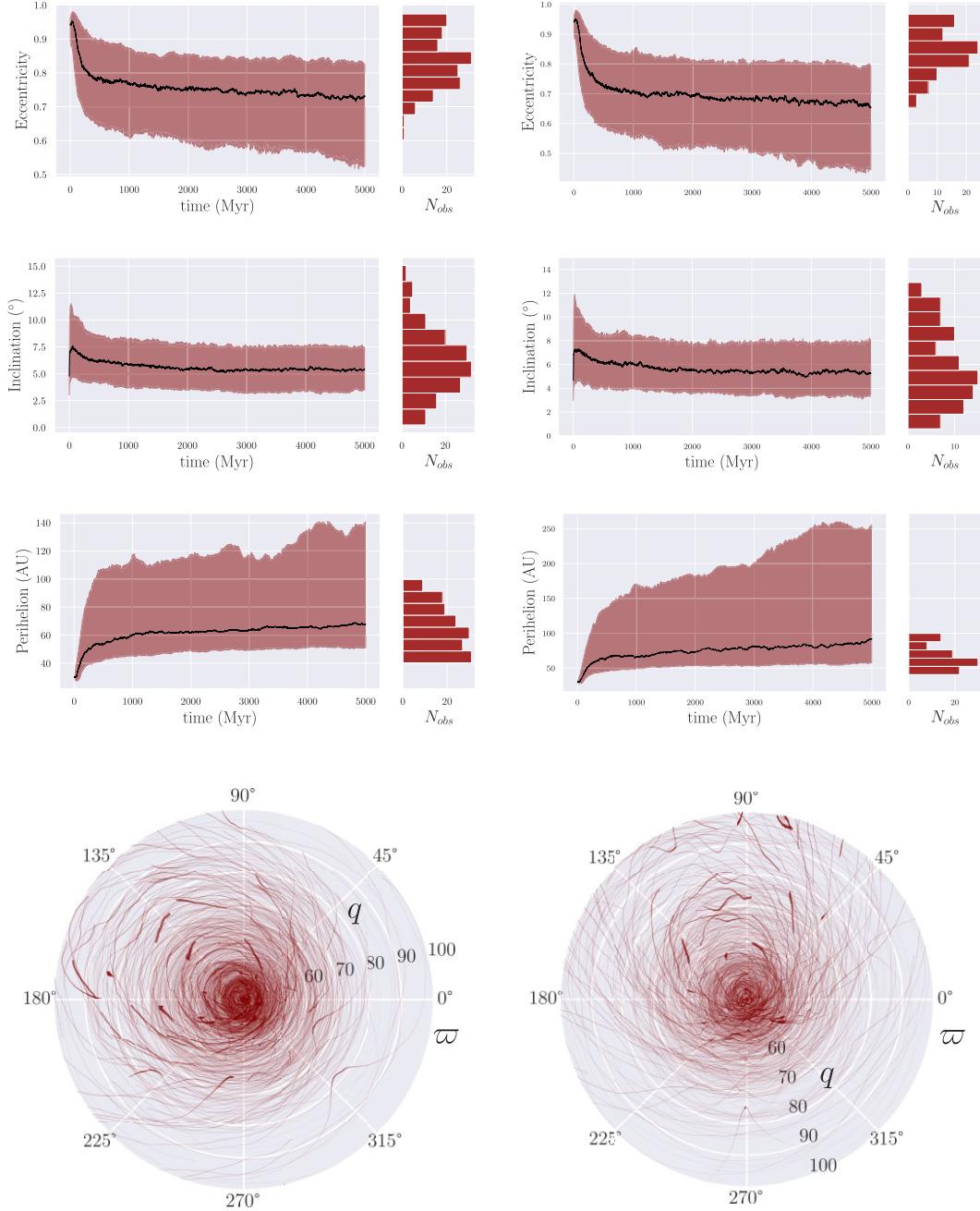

**Figure C2.** Long-term evolution of a $30\,\mathrm{M}_\oplus$ (left column) and $40\,\mathrm{M}_\oplus$ (right column) scattered disc with self-consistent resolution of Neptune (and Jupiter, Uranus, and Saturn as $J_2$). Medians are denoted in black for eccentricity (first row), inclination (second row), and perihelion (third row). The regions between the upper and lower quartiles for each quantity are shaded, and terminal distributions at $t = 5\,\mathrm{Gyr}$ are provided on the right (when filtered for the observational region). In the fourth row, traces for the final Gyr are presented in $\varpi-q$ space. All signatures of the inclination instability are inhibited, and no $\varpi$ bins show grouping (similar to the $20\,\mathrm{M}_\oplus$ case). Inclination instability is suppressed, and $q$ broadening occurs through a distinct mechanism (outlined in Section 3.3).

This paper has been typeset from a TeX/LaTeX file prepared by the author.